\documentclass[amsmath, amssymb,showpacs,aps,prb,reprint,superscriptaddress,floatfix]{revtex4-1}
\usepackage{graphicx} 
\usepackage{dcolumn}
\usepackage{bm}
\usepackage[applemac]{inputenc}
\usepackage{xcolor}
\usepackage{caption}

 \usepackage{hyperref}
\usepackage{enumerate}
\begin{document}


\title{``Butterfly Effect" in Shear-Banding Mediated Plasticity of Metallic Glasses}


 \author{Baoan Sun}
 \thanks{The authors contributed to the paper equally.}
\affiliation{Institute of Physics, Chinese Academy of Sciences, Beijing 100190, China}
\affiliation{Songshan Lake Materials Laboratory, Dongguan, Guangdong 523808, China}

\author{Liping Yu}
\thanks{The authors contributed to the paper equally.}
\affiliation{College of Science, Henan University of Technology, Zhengzhou 450001, China}


\author{Gang Wang}
\affiliation{Laboratory for Microstructures, Institute of Materials, Shanghai University, Shanghai 200444, China}
\author{Xing Tong}
\affiliation{Laboratory for Microstructures, Institute of Materials, Shanghai University, Shanghai 200444, China}
\author{Chuan Geng}
\affiliation{Laboratory for Microstructures, Institute of Materials, Shanghai University, Shanghai 200444, China}

\author{Jingtao Wang}
\affiliation{Herbert Gleiter Institute For Nanoscience, Nanjing University of Science and Technology, Nanjing 210094, China}

\author{Jingli Ren}
\email{renjl@zzu.edu.cn (J. L. R), whw@iphy.ac.cn (W. H. W)}
\affiliation{School of Mathematics and Statistics, Zhengzhou University,
 Zhengzhou 450001, China}

\author{Weihua Wang}
\email{renjl@zzu.edu.cn (J. L. R), whw@iphy.ac.cn (W. H. W)}
\affiliation{Institute of Physics, Chinese Academy of Sciences, Beijing 100190, China}
\affiliation{Songshan Lake Materials Laboratory, Dongguan, Guangdong 523808, China}


\date{\today}

\begin{abstract}
The chaotic dynamics describes how a small change of initial conditions can result in a large difference in a  deterministic nonlinear system, i.e. the "butterfly effect".  Through a combination of experimental and theoretical analysis, here we showed unambiguously that the deformation of metallic glasses (MGs) exhibits such effect where the experimentally observed plasticity displays a large plasticity fluctuation under the normally same conditions. The "butterfly effect" for the plasticity of MGs is related to the chaotic dynamics of a shear band, evidenced by the existence of a torus destroyed phase diagram, a positive Lyapunov exponent and a fractional Lyapunov dimension. Physically, the chaotic shear-band dynamics arises from the interplay between structural disordering and temperature rise within the shear band, which could lead to an uncertainty on the appearance of the critical condition for runaway shear banding events. Our results provide a new perspective on the plasticity of MGs from the viewpoint of complex dynamics and are also important for evaluating the plastic deformation properties of MGs in practical applications. 

\end{abstract}


\maketitle
\section{\label{I}Introduction}
Arising from the long-range disordered atomic structure, the plastic flow of metallic glasses (MGs) or other disordered solids is completely different from that of crystalline materials\cite{Schuh20074067, ChenmwAnnuRew}. At atomic scale, the plastic flow of MGs is found to initiate from some loosely packed atomic-scale regions or soft spots\cite{SPAEPEN1977407, ARGON197947, FalkPRE2009}, which are often correlated by long-range elastic interactions and thus can be organized into cascade or avalanche-like events\cite{Vandem,Lematry2004}. After plastic yielding, these local regions could further concentrate plastic strain, ultimately resulting in  strong flow localization, i.e., the plastic strain is highly localized into shear bands with a thickness of a few tens of nanometers\cite{Greer201371,Krisponeit:2014aa}. The shear banding process could have a profound effect on their macroscopic mechanical behavior. One direct consequence is that a shear band tends to become runaway with the work softening during deformation\cite{ChengPhysRevB}, resulting in the catastrophic failure of MGs. As a result, the poor ductility/plasticity of MGs has become one of main obstacles impeding their widespread applications\cite{Hofmann:2008aa,Sarac:2013ab,JDas}. Meanwhile, the plastic shear of MGs also resembles many important phenomena in natural science and engineering, such as lubrication\cite{Thompson792}, friction\cite{Bhushan:1995ab} and earthquakes\cite{Langer30041996}. Despite the fundamental and engineering importance, a comprehensive understanding on the plastic flow process and its correlation with the macroscopic plasticity in MGs are still lacking.

The dynamics of shear banding process has an important effect on the macroscopic plasticity of MGs. Under tension, a shear band will quickly propagate and become unstable under a tensile stress, leading to almost zero plastic deformablity of MGs. However, some MGs could display some plasticity, where a shear band often proceeds in a stable and intermittent manner\cite{Song2008813,Wright,Loffler}. It was found that there is a close correlation between shear-band stability and the overall plasticity in MGs. Extensive studies showed that the instability of a shear band is controlled by a critical parameter, such as a critical shear velocity or a critical elastic energy density released during a serrated event\cite{WU2010157,Sun:2016aa}. The shear-band instability was also shown to depend on various intrinsic material properties and extrinsic experimental factors, e.g. chemical composition\cite{Lewandowski2005}, sample size\cite{Jang2010ab, KE2014180}, loading rate and even testing machine stiffness\cite{Han20091367, Sun2013PRL}, which ultimately affect the plasticity of MGs. To account for these effects, a number of phenological theories or criteria were proposed, such as the Poisson's ratio criterion for the intrinsic plasticity of MGs\cite{Lewandowski2005} and the shear-band instability index\cite{Han20091367, ChengPhysRevB}.

Despite the efforts above, many mysteries regarding the shear-band mediated plasticity of MGs are still puzzling. A particular issue is that a MG often displays large fluctuation on the plasticity even tested at the same conditions, i.e., chemical composition, sample size and testing conditions. For example, it was shown\cite{yu2009}that the compressive plastic strain of a typical Zr-based MG could vary from less than 5\% to more than 30\% when a large number of specimens were tested under the same conditions. The large variability of plasticity in MGs is generally attributed to the slight change of internal structural states (such as the free volume content), which are inhibited from one of possible structural configurations (inherent states) during the liquid quenching\cite{stillingernature}. Yet, the mechanism underlying that why a small variation of internal states could cause a so large change of macroscopic plasticity in MGs is poorly understood.

Since first proposed in 1960s, chaotic dynamics has been widely observed in many complex systems including weather and climates\cite{Lorenz1963}, geology\cite{PASTERNACK1999253}, biology\cite{Eduardo2012} and computer science\cite{AKHAVAN20111797}. Within the apparent randomness, the chaos describes that how a small change of initial conditions can result in a large difference in a deterministic nonlinear system\cite{Werndl2009}. Here, we understand the sensitivity of the plasticity on the initial conditions of MGs from the viewpoint of complex shear-band dynamics. Through the combined experimental and theoretical analysis, we showed that the dynamics of a single shear band displays the typical characteristics of chaos. The chaotic shear band dynamics could lead to the uncertainty for the appearance of the critical condition for runaway shear banding during deformation, resulting in a large fluctuation on the plasticity of MGs under the same conditions. The physical origin for the chaotic shear-band dynamics as well as its implications for understanding complex nature of amorphous materials are also discussed.

\section{\label{II}Results}

\subsection{The plasticity fluctuation of MGs}
We first show experimentally that the large plasticity fluctuation exhibited by the MGs samples with the same composition and testing conditions. To do this, we performed the compression test on a typical MG with the composition of Zr$_{52.5}$Cu$_{17.9}$Al$_{10}$Ni$_{14.6}$Ti$_{5}$\\(Vit105), which showed a quasi-brittle deformation behavior according to previous studies\cite{Sun2010}. Here, a total of 100 bulk specimens were tested at the same conditions, i.e., the same sample shape and size (rods with a diameter of 2 mm and an aspect ratio of 2:1) and the same testing rate (a constant strain rate of $5\times10^{-4}$ $s^{-1}$). Besides, these bulk MG specimens were also obtained under the same casting conditions. Special cares were also taken during the sample preparation for compression tests, ensuring the same level of surface roughness on both sample ends as well as the parallelism of the two ends for all test specimens (See Figure S1 and S2 in Supplementary Information (SI)). Figure \ref{fig1}(a) shows typical results of compressive stress-strain curves. One can see that even tested at the same conditions, these specimens still display a large variation on the plasticity, from less than 2 \% to more than 20 \%.  A simple statistics showed that most specimens exhibit the plasticity in the range of 5-15\%, and a few specimens display the plasticity of less than 2 \% or more than 25 \%. During the test, most specimens are found to deform by forming a dominant shear band about $45^{\circ}$ inclined to the loading axis, while there are also a few ductile specimens display many multiple shear bands. However, even for those ductile specimens, a primary shear band can still be clearly seen and the final fracture of these specimens is still along the primary shear band(see Figure S3 in SI). The case is different from some ductile MGs where multiple shear bands could form simultaneously at initial  plastic deformation stage after yielding, and the deformed sample finally becomes a "barrel" shape\cite{Sun2010,Sarmah20114482,Liu1385}. All these evidences indicate that the plasticity is mainly determined by the instability of a single dominant shear band for the Vit105 MG\cite{Song2008813, Sun2010}.

We further analyzed the statistics on these plastic strain values. Figure \ref{fig1}(b) shows the probability distribution on plasticity values. Although the number histograms display a typical peak distribution with the plasticity concentrated on a range of values, they seem not follow the Gauss distribution. This indicates that the fluctuation of plasticity is not caused by small stochastic errors of testing conditions, but reflects some hidden physics on the plasticity of MGs.  We also analyzed the cumulative probability distribution of these plastic strain values, $f(x)=P(\varepsilon_{p}<x)=n/N$, where $n$ is the number of specimens with the plasticity $\varepsilon_{p}$ smaller than a given strain value, $x$, and $N$ is the total number of specimens. As can be seen, the $f(x)$ concerning all MG specimens is well fitted by the Weibull distribution, $f(x)=1-exp[-(x/\beta)^{m}]$, where $m$ is Weibull modulus and $\beta$ is a parameter. The Weibull distribution is often used to evaluate the statistic stability of mechanical properties of a material\cite{Courtney, Wu2008}. A small value of Weibull modulus, $m$, often indicates a large variability of the material property, and vice versa. Here, the  values of $m$ by fitting the plasticity data is about 1.05, which is lower than that reported for other MGs in previous studies\cite{yu2009}. The low value of $m$ found here indicates the significant variability on the shear-band instability for the present MG. We also performed the statistics on the exothermic enthalpy of the sub-$T_{g}$ relaxation, $\Delta H_{e}$, and the yield strength, $\sigma_{y}$ over these specimens (For details see Figure S4 and Figure S5 in SI). One can see that the distribution of both $\Delta H_{e}$ and $\sigma_{y}$ are well fitted by the Weibull function. The value of Weibull modulus, $m$, by fitting the $\Delta H_{e}$ data for the current Vit105 MG is 5.42, which is very close to that of Zr$_{65}$Cu$_{15}$Ni$_{10}$Al$_{10}$ MG ($m\sim4.1$) reported by Yu et al\cite{yu2009}. The fitting value of $m$ for the $\sigma_{y}$ data of Vit105 MG is 24.2, which is smaller than that of Zr$_{65}$Cu$_{15}$Ni$_{10}$Al$_{10}$ MG ($m\sim 41.2$), but is comparable to that of the brittle (Zr$_{48}$Cu$_{45}$Al$_{7}$)$_{98}$Y$_{2}$ MG \cite{Wu2008}. Both $\Delta H_{e}$ and $\sigma_{y}$ are related to the free volume content in the as-cast samples\cite{Sun2015211}. The value of $m$ for $\Delta H_{e}$ and $\sigma_{y}$ are much larger than that of plasticity, indicating that the high uniformity of the initial free volume contents over as-cast MG samples.

\begin{figure*}
 \centering
\includegraphics[width=15cm]{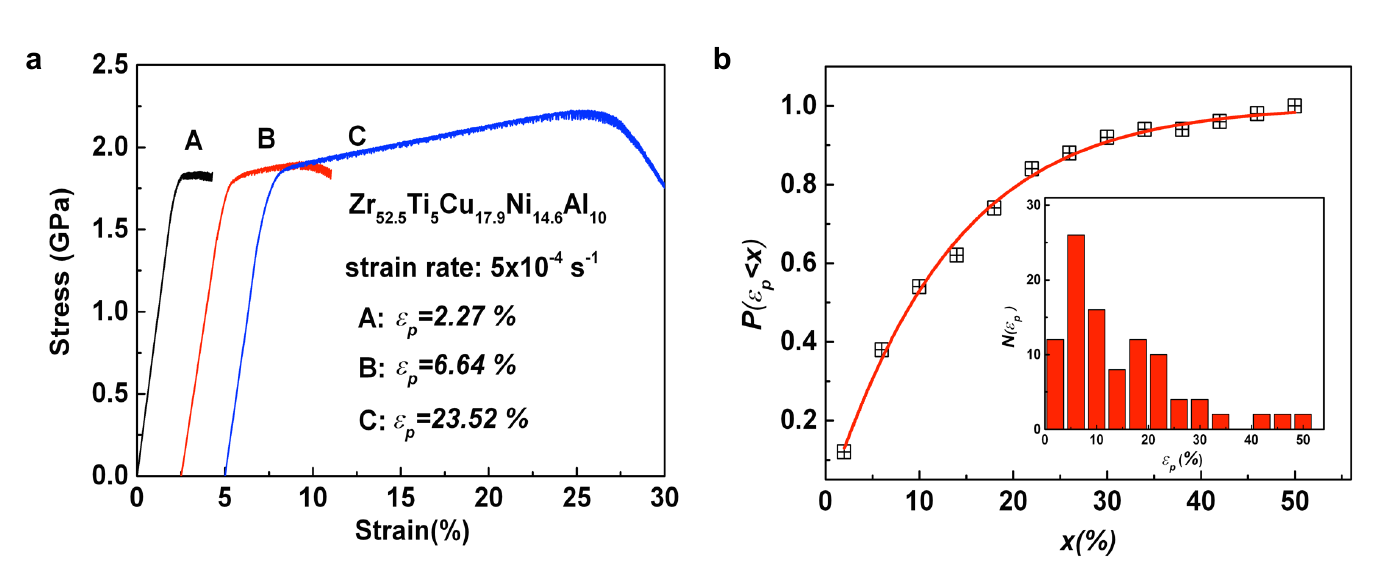}
\caption{\textbf{The plasticity fluctuation of a typical bulk MG (Zr$_{52.5}$Cu$_{17.9}$Al$_{10}$Ni $_{14.6}$Ti$_{5}$) under compression tests.} (a) Three typical stress-strain curves of the MG tested at the same conditions, showing final plastic strain ranging from about 2 \% to more than 20 \%. (b) The cumulative probability distribution of plasticity values for 100 MG specimens, which can be well fitted by the Weibull distribution. The inset shows the counted number histograms for these plasticity values.}\label{fig1}
\end{figure*}

 \subsection{Time series analysis on stress-time curves}
\begin{figure*}
 \centering
\includegraphics[width=15cm]{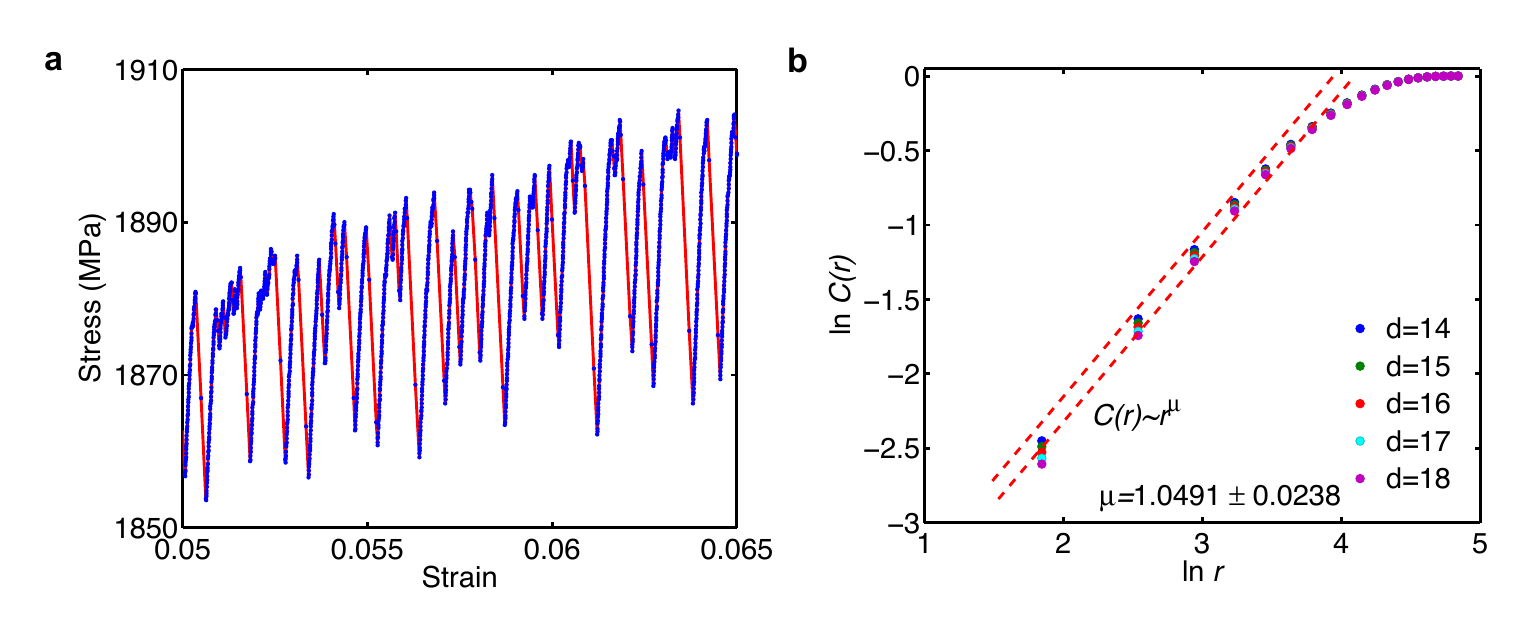}
\caption{\textbf{Time series analysis on the stress-time/strain signals of Vit105 MG.} (a) A typical stress-strain segment for time series analysis (b) The calculated correlation integral $lnC(r)$ versus the distance $r$, plotting in the log-log form, for the embedding dimension $d=14-18$. The data points can be well fitted by the equation $C(r)\sim r^{\mu}$ with the fitting exponent $\mu=1.05$.}\label{fig2}
\end{figure*}

 In the plastic deformation regime, the MG exhibits obvious serrated flow behavior after yielding, as can be seen in enlarged segments of stress-time/strain curves (Figure \ref{fig2}(a)). The serrations are characterized by repeated cycles of sudden stress drops followed by slow upward elastic loading. As extensively studied in literature\cite{Song2008813, mass20113205, Sun2010}, serrated flow is closely correlated with the intermittent shear banding process in MGs. Especially for those MGs deformed by a single dominant shear band, serrated flow has been shown to arise from the stick-slip motion of the band along the primary shear plane\cite{Sun2013PRL}. To uncover the underlying shear band dynamics, we performed nonlinear time series analysis on serrated stress-time signals. The method is particularly useful to extract the hidden dynamic informations of a system from its irregular time noise series. Here, we calculated two parameters: the correlation dimension and the Lyapunov exponent, which are often used to quantify the strange attractor with self-similar properties and the sensitivity to initial conditions of the system\cite{Sarmah20114482}, respectively. Given a scalar time series measured in units of sampling time $\delta t$[$x(k), k=1,2,3,...N$], one can construct $d$-dimensional vectors: $Y_{k}=[x(k), x(k+\tau), ..., x(k+(d-1)\tau)]$, where $\tau$ is the delaying time and can be obtained from the autocorrelation time or from mutual information\cite{PhysRevA331134}. The correlation integral is calculated as \cite{Grassber1983189} (see Methods for details):

  \begin{equation}\label{eq1}
C(r)=\frac{1}{N_{p}}\sum_{i,j}\Theta(r-\mid Y_{i}-Y_{j}\mid)
                          \end {equation}

 where $\Theta$ is the step function and $N_{p}$ is the number of vector pairs summed. For a self-similar attractor $C(r)\sim r^{\mu}$ in the limit of small $r$, where $\mu$ is the correlation dimension. The Lyapunov exponents are calculated by the Wolf's method\cite{WOLF1985285}:
    \begin{equation}\label{eq2}
\lambda_{1}=\frac{1}{t_{M}-t_{0}}\sum^{M}_{k=1} \ln \frac{L^{'}(t_{k})}{L(t_{k-1})}
 \end {equation}
 Here, $M$ is the total number of repeated steps. $L^{'}(t_{k})$ and $L(t_{k-1})$ are two distances defined in Methods. For dynamical system, the spectrum of Lyapunov exponents can be calculated by Benettin algorithm\cite{Benettin1980}, which is equivalent to Wolf's method. Then Lyapunov dimension can be obtained by\cite{FREDERICKSON1983185},
 \begin{equation}\label{eq3}
D_{L}=j+\displaystyle\frac{\sum_{i=1}^{i=j}\lambda_{i}}{|\lambda_{j+1}|}
 \end {equation}
 where $j$ is the largest inter such that $\sum_{i=1}^{i=j}\lambda_{i}\geq 0$.

With equations above, we analyzed a typical serrated stress-time curve of Vit105 MG. Figure \ref{fig2}(b) displays the variation of correlation integral, $C(r)$ with the distance, $r$,  for different embedding dimensions. It can be seen that the slope of $lnC(r)$ versus $lnr$ converge to 1.05 as the embedding dimension reaches $d=14$. Thus the correlation dimension is taken to be $\mu=1.05$, indicating the self-similar strange attractor for time-stress signals. In addition, the calculated Lyapunov spectrum also exhibits a  positive maximum exponents ($\lambda_{1}=0.0321$). The existence of a finite correlation dimension and a stable positive Lyapunov exponent strongly suggests that the serrated stress signal has chaotic dynamics.

 \subsection{Theoretical modelling and analysis}
     We attempt to understand the above results from the complex dynamics of a single shear band. This is consistent with the experimental facts that the deformation of most MGs are mainly dominated by a single primary shear band during compression\cite{Song2008813, mass20113205}. The first step is to construct a set of dynamic equations of the shear band, which can fully describe its motion and internal evolution. Previous studies\cite{Sun2013PRL, Daub2013} have proposed a stick-slip dynamic model to explain the origin of serrated flow in MGs. The model considers that the elastic energy released in the sample-machine system as the driving force for shear-band motion, while the plastic energy is mainly dissipated as configurational entropy or structural disordering within shear band. The dynamic equations in the model are written as\cite{Sun2013PRL}:
\begin{equation}\label{eq4}
\frac{d\sigma_{b}}{dt}=k(v_{0}-\lambda_{b}\dot{\epsilon_{b}})
\end {equation}
\begin{equation}\label{eq5}
\frac{d\chi}{dt}=\frac{\dot{\epsilon_{b}}\sigma_{b}}{c_{0}}\left[1-\frac{\chi ln(\dot{\epsilon_{c}}/\dot{\epsilon_{b}})}{\chi_{w}} \right]
\end {equation}
where $v_0$ is the loading rate, $k$ is the effective elastic constant of sample-machine system, $\chi$ is the effective disorder temperature, $\sigma_{b}$ and $\dot{\epsilon_{b}}$ are the stress resistance and strain rate in the shear band, respectively. $\chi$ is the effective disorder temperature\cite{Daub2013}, which describes the internal state evolution within shear band. $\chi$ could reach a stable state $\chi_{w}/ln(\dot{\epsilon_{c}}/\dot{\epsilon_{b}})$ for a constant $\dot{\epsilon_{b}}$. The definition of other parameters can be found in Ref. [24]. The constitutive relation between $\dot{\epsilon_{b}}$ and $\sigma_{b}$ is given by the cooperative shear model (CSM)\cite{Johnson2006}: $\dot{\epsilon_{b}}=\dot{\epsilon_{s}}exp (-1/\chi)exp[-W_{0}(\sigma_{b0}-\sigma_{b})^{(3/2)}/(k_{B}T)]$, where $W_{0}$ and $\sigma_{b0}$ are the critical energy barrier for shear transformation zones (STZs) and the yield strength of the glass at 0 K, respectively. The term $exp(-1/\chi)$ is proportional to the number density of STZs.

   The above model could predict the appearance of serrated flow for shear banding, which is related to a critical stiffness parameter. However, the serrations calculated by this model only have a periodic solution with a fixed size. To fully reveal the complex shear-band dynamic, here we extend the above model by considering the temperature evolution within the shear band. In principle, there should be some temperature gradient within a shear band from its center to boundaries. Here, we assume shear band as a thin layer with homogeneous temperature distribution. In this case, the flow of heat energy (per unit time and per unit area) across the shear-band boundary is proportional to $K\Delta T/a$, where $\Delta T=T-T_{R}$ is the instantaneous temperature difference between shear band (with a temperature $T$) and glassy matrix (with room temperature $T_{R}$), $K$ is thermal conductivity and $a$ is a characteristic length with the same order of magnitude as shear band thickness. Meanwhile, the heat energy is also produced within the band by the work done by plastic deformation. The dynamic equation for temperature evolution reads:

 \begin{equation}\label{eq6}
 \frac{d\Delta T}{dt}=\frac{\alpha \sigma_{b}\dot{\epsilon_{b}}}{\rho c_{p}}-\frac{K\Delta T}{a}
\end {equation}
where $\rho$ and ${c_{p}}$ are the density and the heat capacity of the glass material, respectively, $\alpha$ is the fraction of plastic work dissipated as heat.

\begin{figure*}
 \centering
\includegraphics[width=15cm]{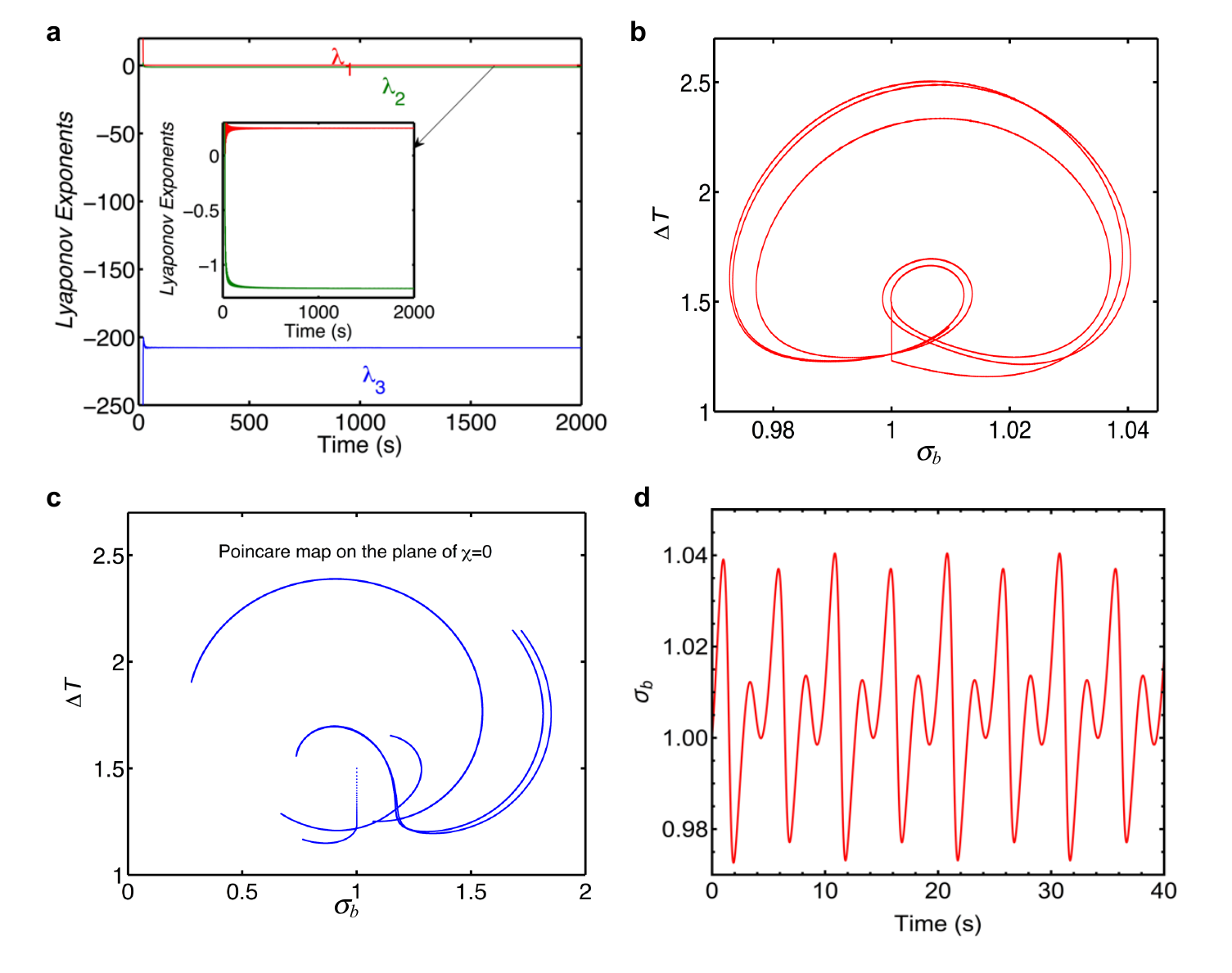}
\caption{ \textbf{Chaotic dynamics of a single shear band.} (a)  Plot of Lyapunov exponent spectrum. The inset is the local amplification for the spectrum. The largest Lyapunov exponent $\lambda_{1}=0.2489$. (b) The phase diagram in the plane of $\sigma_{b}-\Delta T$, showing a self-similar destroyed torus. (c) Poincar\'e map on the plane of $\chi=0$ with an irregular invariant. (d) Time history diagram of $\sigma_{b}$ calculated under the initial condition (1, 0.4, 1.5). All results are calculated at the same parameter values: $k=2000$, $\dot{\epsilon_{0}}=10^{4}, \dot{\epsilon_{s}}=10^{6}$, $\dot{\epsilon_{c}}=10^{5}$ and $\beta=10^{4}$.}\label{fig3}
\end{figure*}

      Eqs.\ \eqref{eq4}-\eqref{eq6} describe the dynamics of a single shear band under compression. Here, $\chi$ and $\Delta T$ are two internal state variables, $k$ and $v_{0}$ are two parameters which can be adjusted, while other parameters are set as constants. We perform a thorough theoretical analysis and numerical calculations on the dynamic model. For analysis details, one can see Text SI in the SI. The results show that the shear band dynamic exhibits a chaotic state within some range of $k$ and $v_{0}$. A typical example for $k=2000$ and $\dot{\epsilon_{0}}=10^{4}$ ($\dot{\epsilon_{0}}$ is defined by $\dot{\epsilon_{0}}=v_{0}/\lambda$) is shown in Figure \ref{fig3}. As can be seen, the self-similarity of attractor and sensitivity to initial conditions of the chaotic state are characterized by a positive largest Lyapunov exponent ($\lambda_{1}=0.2489$) and Lyapunov dimension ($D_{L}=1.2047$). The self-similar torus destroyed phase diagram and the irregular invariant of Poinca\'{e} map (Figure \ref{fig3}(b) and (c)) also imply a chaotic state.  Arising from the chaotic dynamics, the history diagrams of $\sigma_{b}$ with time display serrations with multiple magnitudes of stress drops, as shown in Figure \ref{fig3}(d). These characteristics are very similar to those of hopping type-B bands in Portevin-Le Chatelier (PLC) effect\cite{SARMAH2015192}. It is worth noting that with the chaotic dynamics was also reported by Lema\^{\i}tre in the boundary lubrication by using a full generalization of STZ theory as the constitutive equations\cite{LematrePRE2004}, which is similar to the stick-slip shear band dynamics observed here.

\begin{figure*}
 \centering
\includegraphics[width=15cm]{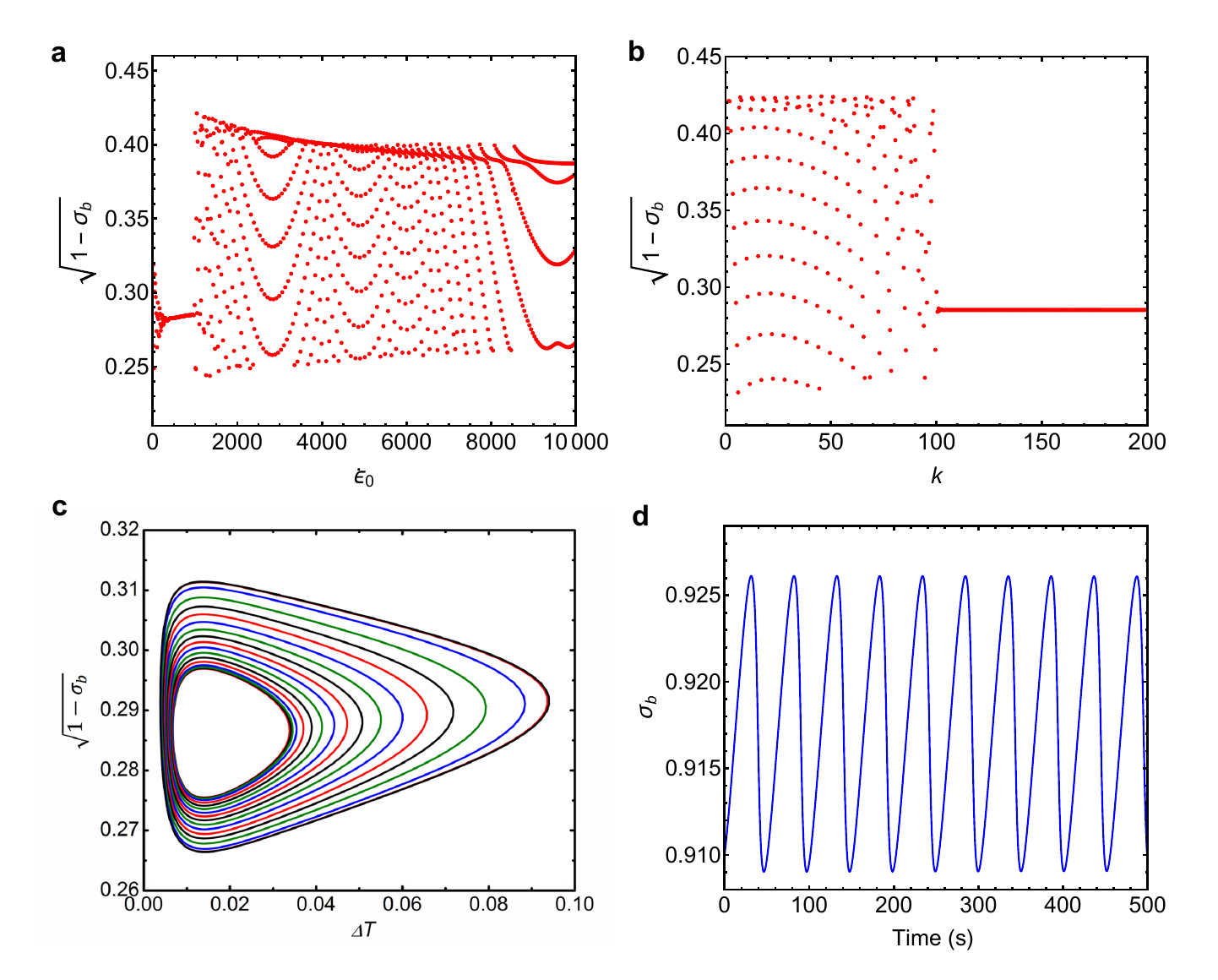}
\caption{ \textbf{Periodic dynamics of a single shear band} (a) The calculated bifurcation diagram of the model with $\dot{\epsilon_{0}}$ in the range of $[1, 10^{4}]$ for $k=100$, $\dot{\epsilon_{s}}=10^{6}$,
$\dot{\epsilon_{c}}=10^{5}$ and $\beta=10^{5}$ under initial condition (0.3, 0.2, 0.013). (b) The calculated bifurcation diagram of the model for $\dot{\epsilon_{0}}=10^{3}$, $\dot{\epsilon_{s}}=10^{6}$,
$\dot{\epsilon_{c}}=10^{5}$ and $\beta=10^{5}$ under initial condition (0.3, 0.2, 0.013).  From these diagrams, one can see that Hopf bifurcation is initially subcritical and then become supercritical with increasing value of
$\dot{\epsilon_{0}}$,  while the direction of Hopf bifurcation can not change with varying $k$.  (c) The limit cycles bifurcated by Hopf bifurcation. (d) The time history diagram of $\sigma_{b}$ corresponding to the periodic orbit.}\label{fig4}
\end{figure*}

     In addition to the chaotic dynamic state, the shear band dynamics also has a periodic solution when the external strain rate $\dot{\epsilon_{0}}$ and the elastic constant $k$ are decreased. For the sake of discussion, we replace $\sigma_{b}$ with $u=\sqrt{1-\sigma_{b}}$ as a new variable, while keep $\chi$ and $\Delta T$ unchanged. From the theoretical analysis shown in Text SI in SI, ($u$, $\chi$, $\Delta T$) has a positive equilibrium point $E_{3}$. For $\dot{\epsilon_{0}}=10^{3}$ and $k=100$,  $E_{3}$ has a value (0.3256, 0.2063, 0.01341) with characteristic roots $\mu_{1}=-9.18\times10^{4}$ and $\mu_{2,3}=\pm 0.128i$, indicating Hopf bifurcation emerging at $E_{3}$. There are two special points where degenerate Hopf bifurcation occurs: $k=92.69$, $\dot{\epsilon_{0}}=202.35$ and $k=1000$, $\dot{\epsilon_{0}}=0$. Figure \ref{fig4}(a) and (b) display the bifurcation diagram of the model with $k$ and $\dot{\epsilon_{0}}$, respectively. From these diagrams, one can see that the supercritical Hopf bifurcation arises at the point $k=100$, $\dot{\epsilon_{0}}=1415.8$. The Hopf bifurcation is initially subcritical and then becomes supercritical with increasing value of $\dot{\epsilon_{0}}$,  while the direction of Hopf bifurcation cannot change with varying $k$. In nonlinear dynamics, Hopf bifurcation is often associated with the emergence of limit cycle. Figure \ref{fig4}(c) displays limit cycles initiated from Hopf bifurcation for values of $k$ and $\dot{\epsilon_{0}}$  in the present model. The existence of limit cycles clearly suggests the periodic oscillation in the system. The periodic solution can also be verified by the numerically calculated time profile of $\sigma_{b}$, which shows periodic stress serrations with a fixed size (Figure \ref{fig4}(d)).

  \section{\label{III}Discussion}
    From experimental and theoretical analysis above, one can see that a shear band could exhibit complex nonlinear dynamics including chaos and periodic orbit. These dynamics behavior should have a profound effect on the shear-band stability and the ultimate plasticity of MGs. Now let's discuss the correlation between the chaotic shear-band dynamics and the plasticity fluctuation observed in experiments. Previous studies showed that the stability of a single shear band can be related to a critical parameter, such as a critical shear velocity \cite{Sun:2016aa}or a critical elastic energy density\cite{WU2010157} released during a serrated event. Once a shear band reaches the state set by the critical parameter, a runaway ``defect" or crack will be developed within the band, resulting in the catastrophic materials failure. The plastic strain at which the critical state appears within the shear band defines the macroscopic plasticity of MGs. Since shear banding process in compression often proceeds in a stick-slip manner, so the critical state for shear band can be associated with a critical serrated event. According to our present analysis, if the shear band dynamics is chaotic, the appearance of the critical state for the shear-band instability will be sensitive to initial conditions. Practically, there will be tiny differences (e.g. on chemical compositions, free volume contents, sample size,  etc) due to experimental errors during sample preparation and testing process. These tiny differences will lead to slight different initial conditions for the shear banding, and eventually are enlarged through the inherent chaotic dynamics. As a result, the appearance for the critical state at which shear band become runaway is unpredictable during deformation, resulting in the large plasticity fluctuation of MGs, as observed in experiments. The chaotic dynamic behavior of shear band found in MGs is reminiscent of the ``butterfly effect" as observed in many complex dynamic systems\cite{Werndl2009}, e.g. a butterfly flapping its wings in Brazil can cause a hurricane in Texas .
    
 \begin{figure*}
 \centering
\includegraphics[width=15cm]{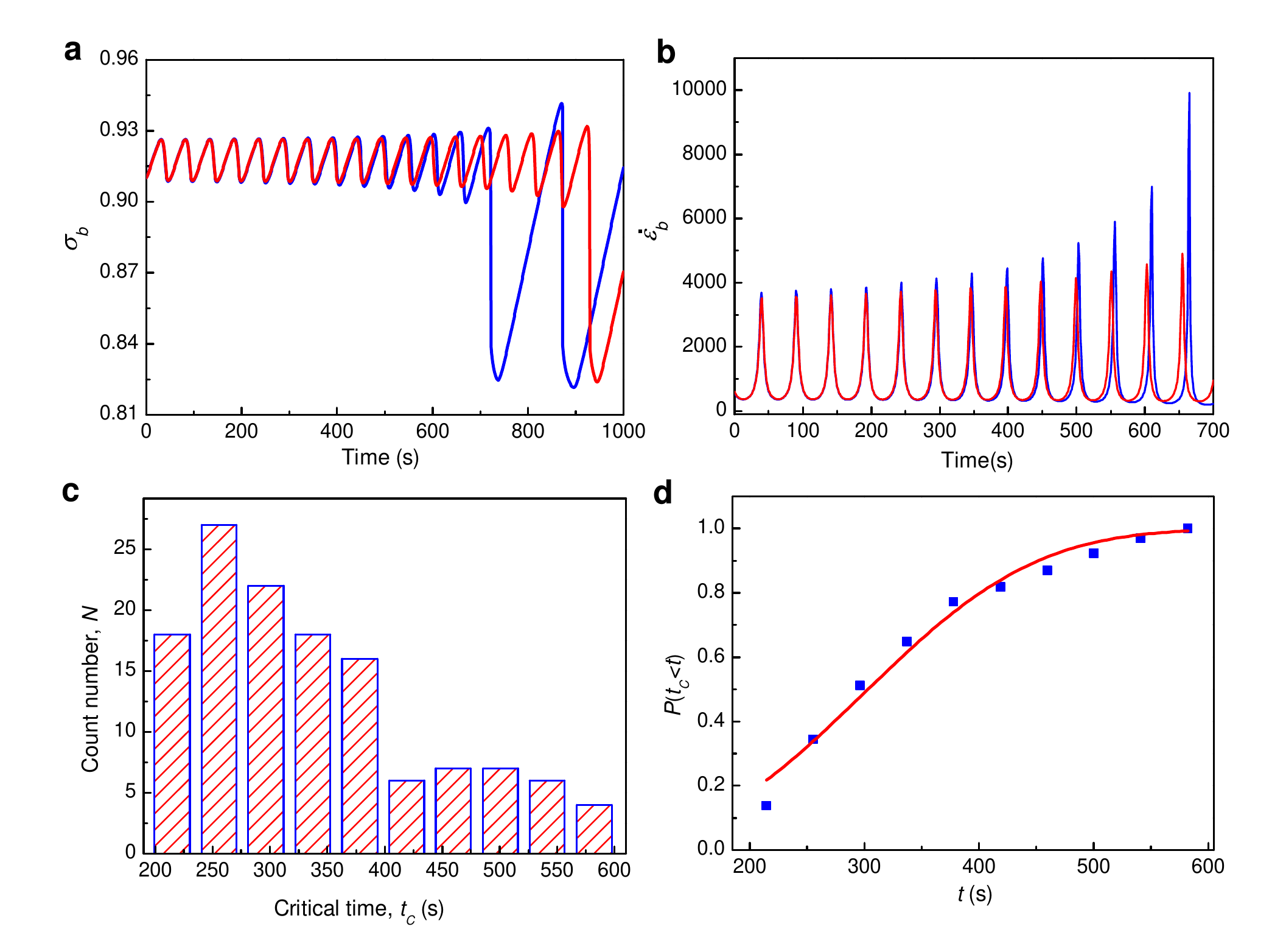}
\caption{\textbf{Numerical calculations illustrating the sensitivity of shear band properties on the initial condition in the chaotic dynamic regime for $k=100$ and $\dot{\epsilon_{0}}=10^{3}$.} (a) Time evolution of resistant stress $\sigma_{b}$ of a shear band under the initial condition (0.91, 0.1975, 0.013) (in the blue line) and (0.91, 0.2, 0.013)(in the red line), respectively. (b) Time evolution of the strain rate within a shear band,$\dot{\epsilon_{b}}$, corresponding to (a). (c) The count number histograms for critical time, $t_{c}$, when  $\dot{\epsilon_{b}}$ reaches 4500 for the first time for 133 parameter sets of initial conditions from (0.90901, 0.1975, 0.013) to (0.91,0.1975,0.013) with an interval of 0.00001 and from (0.91,0.197,0.013) to (0.91,0.2,0.013) with an interval of 0.001.  (d) The cumulative probability distribution of $t_{c}$ for 133 parameter sets of initial conditions, which can be well fitted by the Weibull distribution with the fitted parameter $\beta=342.48$ and $m=3.01$.}\label{fig5}
\end{figure*}

   To further illustrate the ``butterfly" effect in the shear-band mediated plasticity in MGs, we numerically calculated the evolution of various variables of shear band with time in the chaotic dynamics regime, as shown in Figure \ref{fig5}. As can be seen, when the initial conditions of ($\sigma_{b}$,$\chi$,$\Delta T$) is slightly changed from (0.91, 0.1975, 0.013) to (0.91, 0.2, 0.013), the evolution of $\dot{\epsilon_{b}}$ within shear band can be significantly changed after $t= 600 s$. Particularly, the serrated event with the largest value of 
   $\dot{\epsilon_{b}}$ which may cause the instability of shear banding occurs at different times. By arbitrary choosing the critical strain rate for the shear-band instability as   $\dot{\epsilon_{bc}}=4000$, we made statistics on the critical time, $t_{c}$, for the occurrence of shear-band runaway event over a large number of parameter sets of initial conditions.  The probability distribution of $tc$ follows the Weibull distribution well (see Figure \ref{fig5}c and d), in accordance with experimental results. The numerical calculations further confirm the deterministic nature of the complex shear band dynamics. As the elastic constant $k$ and the external strain rate $\dot{\epsilon_{0}}$ are deceased, the chaotic dynamics of shear band is gradually transformed to the periodic orbit. The periodic state is not so sensitive to initial conditions, and should correspond to the smaller plasticity fluctuation for the same MG. It should be noted that the dynamic states (chaos and periodic orbit) are only limited to the case of a single dominant shear band. For some special ductile MGs, they could form a large number of multiple shear band simultaneously at the initial plastic deformation stage, and the plasticity is often related to a self-organized critical dynamics arising from the interaction of a large number of multiple shear bands\cite{Sun2010, Sarmah20114482}.

   In general, the simple chaos requires at least three variables in a dynamic system\cite{Lorenz1963}.  In previous stick-slip models, the effective disorder temperature is considered as the only state  governing the internal evolution of shear band. In the cases, serrations formed by stick-slip shear banding show a periodic behavior\cite{Sun2013PRL} rather than a chaotic behavior. In our model, in addition to the effective disorder temperature, we also consider the temperature rise as a internal state variable.The effective disorder temperature is a measure of the STZ density during deformation\cite{JSLanger}, and in fact is a reflection of the structure disordering in MGs. Thus, the interplay between structural disordering and the temperature rise during deformation must play an important role in the formation of chaotic shear band dynamics. In literature\cite{Schuh20074067, Greer201371}, the structural disordering as a main cause for strain softening in MGs is well recognized, yet the temperature rise within shear band has long been debated. By spatially and temporally resolved measurement on shear band velocity, recent studies \cite{Wrigh20094639}showed that the maximum temperature rise during shear band propagation is only a few tens of Kelvins. This temperature rise is not enough to cause the significant softening of glassy materials. Therefore, the thermal effect is often believed to be a consequence of strain softening, rather than its main cause before the final fracture of MGs. The effect of temperature rise on the strain softening is also neglected in many theoretical models for shear localization\cite{SPAEPEN1977407, ARGON197947, FalkPRE2009}. However, our present analysis showed that the role of temperature rise cannot be neglected in shear band dynamics, especially in complicating shear band dynamics. From numerical calculations on our model, we can see that for the periodic orbit state, the calculated temperature rise, $\Delta T$, is small, in the range of (0.01-0.15) $T_{R}$, while for the chaotic dynamic state, $\Delta T$ could reach 2.5 $T_{R}$, a value sufficient to cause the runaway instability of shear band.The significant temperature rise during the shear-band instability event  is also evidenced by the vein-like patterns across the whole surface area and some solidified liquid droplets and wires among them(See Figure S6 in SI). The large temperature rise clearly results from the "butterfly effect" of  shear band dynamics. In the chaotic dynamic regime, the small initial temperature rise can be amplified through interplaying with the internal stress and the structural disordering in shear band.  Conversely, the temperature rise also modulate the other two variables ($\sigma_{b}$ and $\chi$) during deformation, making the shear band dynamics become complex. The current finding suggests that the conventional view on the role of temperature rise during the plastic flow of MGs should be reassessed.

   In summary,  we showed that a single shear band in MGs exhibits complex chaotic dynamics through the combination of experimental and theoretical analysis.  We also demonstrated that the experimentally observed large plasticity fluctuation of MGs tested at the same conditions can be interpreted from the chaotic shear-band dynamics, which could lead to an uncertainty on the appearance of the critical condition for runaway shear banding during deformation. Physically, the chaotic shear-band dynamics arises from the interplay between structural disordering and temperature rise with in the shear band. By tuning the deformation parameters, the chaotic dynamics can be transformed to a periodic orbit state.  Our results suggest that the plastic flow of MGs is a complex dynamic process, which is highly sensitive to initial conditions and reminiscent of the "butterfly effect" as observed in many complex dynamic systems. Finally, it is worth noting that shear banding process in MGs resembles many dynamic phenomena across different time and length scales, e.g. peeling of adhesive tapes\cite{Sarmah20114482}, friction and lubrication in industrial process\cite{Urbakh:2004aa}, seismic and geodetic fault slip\cite{Carlson1989}, \textit{etc}. Our current finding may be useful to understanding the complex nature of these dynamic systems, all of which one may encounter in the broad discipline of nature and engineering.

 \section*{\label{BI} Methods}
\textbf{Sample preparation}: MG alloy ingots with the nominal compositions Zr$_{52.5}$Cu$_{17.9}$Al$_{10}$Ni\\
$_{14.6}$Ti$_{5}$(Vitreloy105) were produced by arc melting a mixture of pure metals (purity $\ge$99.5\% in mass weight) in a Ti-gettered argon atmosphere. To ensure chemical homogeneity, each ingot was remelted for at least three times. Glassy alloy rods with diameters of 2 mm and a length of at least 30 mm were obtained by suction casting into a water-cooling copper mould. To make sure that all glassy samples have the same internal state, the melting current and time during suck casting process for all samples are kept as 130 A and 25 s, respectively.
The amorphous nature of specimens in the as-cast state were confirmed by the x-ray diffraction (XRD) method using a MAC Mo3 XHF diffractometer with Cu K$\alpha$ radiation and the differential scanning calorimetry (DSC, Perkin Elmer DSC7) as well as a high-resolution transmission electron microscopy (HRTEM, JEM 2010 F operating at 200 KV).

\textbf{Mechanical tests}: The uniaxial compression tests were performed on an universal electromechanical test system (New SANS, MTS) under a constant strain rate $5\times10^{-4}$ s$^{-1}$. Specimens about 4 mm long were cut from BMG rods by means of a diamond saw, and then carefully ground into compression specimens with an aspect ratio of 2:1. Both ends of all specimens were ground carefully with different grades of sandpapers (initially 400 grades for 5 mins, then 600 and No.1000 for 5 mins, respectively, and finally No.3000 for 5 mins). The surface roughness of the two sides of samples were analyzed with a laser scanning confocal microscope (OLYMPUS, OLS4000). The surface evenness of all samples are at the same level with the $R_{a}$ in a range of 10-15 nm, as shown in Figure S1 in SI. Special care was taken to ensure that the two surfaces of test specimens were parallel and orthogonal to the loading axis (See Figure S2 in SI). All test specimens are taken from the bottom of glassy rods to exclude the effect of cooling rate on mechanical test results.The load, displacement and the time are recorded at a frequency of 50 Hz. The sample strain was measured by a contact extensometer attached to the testing machine. Special parts are designed to fix the extensometer with the tungsten carbide plates in order to precisely measure sample strain for compression. After the test,  the morphologies of shear bands as well as fracture surfaces were investigated via a scanning electron microscopy (SEM, Gemini1530).

\textbf{Time series analysis} Given a scalar time series measured in units of sampling time [$\sigma(k), k=1,2,3,...N$], where $N$ is the number of data. One can construct a $d$-dimensional vector: $[{Y_{{t_i}}}(d) = \{ \sigma ({t_i}),\sigma ({t_i} + \tau ),...,\sigma ({t_i} + (d - 1)\tau ),{t_i} = 1,...,[N - (d - 1)\tau ]\}$, where $\tau$ is the delaying time and can be obtained from mutual information\cite{PhysRevA331134}, $d$ is the embedding dimension which can be calculated by the Cao method \cite{CAO199743}. For mutual information method, the mutual information quantity between $\tau$ and $t+\tau$ is
\[{I_\tau } = \sum\limits_{k = 1}^N {P[x (k),x (k + \tau )]} {\log _2}\{ \frac{{P[x (k),x(k + \tau )]}}{{P[x (k)]P[x (k + \tau )]}}\}. \]
$I(\tau)$ is dependent of time delay, $\tau$. When $I(\tau)$ reaches its local minimum value at first time, the corresponding $\tau$ is considered as the delay. Then for real delayed experiment time, $t$, corresponding to time series, is obtained by $t=\tau h$, where $h$ is mean value of time interval between two serration events. Define the distance between two point is $|Y_{i}-Y_{j}|=\max\limits_{1\leq k\leq n}|Y_{ik}-Y_{jk}|$. Two points is called as correlated vector if the distance between them is less than a given positive number $r$. Correlation integral is defined by the percent for
correlated vectors among $N_{p}$($N_{p}=N(N-1)/2$) kinds of pairs is , i.e.,
$$C_{n}(r)=\frac{1}{N_{p}}\sum_{i,j=1}^{N}\Theta(r-|Y_{i}-Y_{j}|) $$
where $\Theta$ is Heaviside step function, $\Theta(x)=0,$  for $x\leq0$ and $\Theta(x)=1,$  for $x>0$. For a self-similar attractor, $\lim \limits_{r\rightarrow0} C_{n}(r) \propto r^{\nu},$ where $\nu$ is called as correlation dimension.

For the calculation of Lyapunov exponent and dimension\cite{WOLF1985285}\cite{FREDERICKSON1983185}, take an initial point, $Y(t_{0})$, and its nearest neighbor point, $Y_{0}(t_{0})$, suppose that the distance between these two points is $L(t_{0})$. Tracking the evolution of these two points, after a time, $t_{1}$, two points evolve to be $Y(t_{1})$ and  $Y_{0}(t_{1})$, and the distance is $L'({t_1}) = |Y({t_1}) - {Y_0}({t_1})|>\omega,$  where $\omega$ is a constant and slightly larger than the minimum distance of each two points in the reconstructed space. Then choose $Y_{1}(t_{1})$, which is the nearest neighbor point of $Y(t_{1})$ in case of the orbits running out, and the corresponding distance is $L(t_{1})$.  Tracking the evolution, we can obtain $L'({t_2})$ similarly. Repeat above process until the end of time series, and suppose $M$ as the total number of repeated steps, a series of $L(t_{i-1})$ and $L'(t_{i+1})$, $i=1,2, \cdots, M$  can be obtained. The Lyapunov exponent and Lyapunov dimension are calculated by Eqs. \eqref{eq2} and \eqref{eq3}, respectively.
\\

  \section*{Acknowledgments}
 B. A. Sun and W. H. Wang acknowledge the support from the National Science Foundation of China (NSFC) (Grant Nos: 51822107, 51671121, 61888102 and 51761135125, 51520105001), National Key Research and Development Plan (Grant No. 2018YFA0703603) and the Natural Science Foundation of Guangdong Province (Grant No. 2019B030302010)  and Strategic Priority Research Program of Chinese Academy of Sciences (XDB30000000) and Key Research Program of Frontier Sciences (QYZDY-SSW-JSC017). L. P. Yu thanks the High-Level Personal Foundation of Henan University of Technology (2018BS027). J. L. Ren thanks the support from the NSFC (Grant No: 11771407) and the Innovative Research Team of Science and Technology in Henan Province (17IRTSTHN007).
\section*{References}
\bibliography{mybibfile.bib}








\end{document}